# Low-intensity light switching of cavity-atom polaritons

Yifu Zhu

Department of Physics, Florida International University, Miami, FL

Abstract

I analyze an all-optical switching scheme in a cavity QED system consisting of multiple three-level atoms confined in a cavity mode. A control laser coupled to the atoms from free space induces quantum interference in the coupled cavity-atom system and performs all optical switching of the cavity-atom polaritons. When the cavity is tuned far away from the atomic resonance, the cavity-atom polariton consists of nearly pure photonic excitation and then the scheme can be used for photon switching at low light intensities.

Cavity quantum electrodynamics (Cavity QED) studies interactions of photons and atoms confined in a cavity mode, and has a variety of applications in quantum physics and optics [1]. The fundamental cavity QED system consists of a single two-level atom coupled to a single cavity mode [2-3]. The first-excited eigen-states of the coupled cavity-atom system consist of two normal modes separated in energy by 2g, commonly referred to as the vacuum Rabi splitting ( $g = \mu\sqrt{\omega_c / 2\hbar\varepsilon_0 V}$ is the coupling coefficient, where $\mu$ is the atomic dipole moment, $\omega_c$ is the frequency of the cavity mode, V is the mode volume, $\hbar$ is the Plank constant, and $\varepsilon_0$ is the vacuum permittivity). The two normal modes consist of coherent superposition of the atomic excitation and photonic excitation, and equivalently can be viewed as polaritons of the coupled cavity-atom system. If N two-level atoms interact collectively with the cavity mode, the coupling coefficient becomes $G = \sqrt{N} g$ and the vacuum Rabi splitting of the normal modes becomes 2G [4-6].

Here we combine studies of cavity QED and nonlinear optics at low light levels, and propose a scheme for performing all-optical switching of the cavity-atom polaritons in a multi-atom cavity QED system with an ultra-weak control laser. All-optical switching at low light levels may have practical applications in quantum electronics and quantum information network, and has been a subject of many recent studies [7-15]. In the proposed switching scheme based on cavity QED, we show that with the control laser coupled to the atoms from free space and tuned in frequency to the polariton (the normal mode) resonance, destructive quantum interference is induced in the polariton excitation and can be used for optical switching of the cavity-atom polaritons. In particular, when the cavity is tuned far away from the atomic resonance, one of the polariton excitations (normal modes) consists of nearly pure photonic excitations, or photons, and the scheme can be used effectively to perform all optical switching of cavity photons.

Consider a composite atom-cavity system consisting of a single mode cavity containing N

identical three-level atoms depicted in Fig. 1 [16-19]. The cavity consists of two mirrors with reflectivity R and separated by a length L. The cavity mode couples the atomic transition |a>-|e> and $\Delta_c = \nu_c - \nu_{ea}$ is the cavity frequency detuning from the atomic transition frequency $\nu_{ea}$. A weak probe laser is coupled into the cavity mode ($\Delta_p = \nu_p - \nu_{ea}$ is the probe frequency detuning.) and at appropriate frequency, excites the cavity-atom polaritons. A weak control laser couples the atoms from open free space and drives the atomic transition |b>-|e> with Rabi frequency $2\Omega$ and a frequency detuning $\Delta = \nu - \nu_{eb}$. The collective atomic operators are $\rho_{ae} = \sum_{i=1}^{N} \rho_{ae}^i$ and $\rho_{be} = \sum_{i=1}^{N} \rho_{be}^i$, and couple the symmetric, Dicke-type atomic states, among which the ground state is represented by $| a > = | a_1 .........a_N >$ and the excited state with one atomic excitation is represented by $| e > = \sum_{j=1}^{N} \frac{1}{\sqrt{N}} | a_1 ......e_j ......a_N >$. Without the weak control laser, the ground state of the cavity-atom system is $| a,0 > = | a > | 0 >$, and the two product states with one excitation quanta are $| a,1 > = | a > | 1 >$ and $| e,0 > = | e > | 0 >$ (|1> and |0> are one-photon and zero photon states of the cavity mode). The collective coupling coefficient of the two product states is $G = \sqrt{N} g$. After diagonalizing the interaction Hamiltonian, the two eigenvalues of the first excited states are derived as $\lambda_{\pm} = \frac{\Delta_c}{2} \pm \sqrt{(\frac{\Delta_c}{2})^2 + g^2 N}$, and the corresponding eigenstates are $| \varphi_{\pm} > = a_{\pm} | e > | 0 > + b_{\pm} | a > | 1 >$ with $a_{\pm} = g \sqrt{N} / \sqrt{g^2 N + (\lambda_{\pm})^2}$ and $b_{\pm} = \lambda_{\pm} / \sqrt{g^2 N + (\lambda_{\pm})^2}$. $| \varphi_+ >$ and $| \varphi_- >$ are separated in energy by $\lambda_+ - \lambda_-$ and are commonly referred to as two normal modes of the coupled cavity-atoms system [4-6], or equivalently, polaritons of the atomic excitation and photonic excitation. When $\Delta_c = 0$, $| \varphi_{\pm} > = \frac{1}{\sqrt{2}} (| e,0 > \pm | a,1 >)$, the two normal modes (polaritons) have equal weights of the atomic excitation and photonic excitation. When $\Delta_c >> \sqrt{N} g$, $|\varphi_+> \approx \frac{g\sqrt{N}}{\Delta_c}|e,0> + |a,1>$ ($\lambda_+ \approx \Delta_c$) and $|\varphi_-> \approx |e,0> - \frac{g\sqrt{N}}{\Delta_c}|a,1>$ ($\lambda_- = -\frac{g^2 N}{\Delta_c}$). That is, $|\varphi_+>$ contains nearly pure photonic excitation (a single photon) and $|\varphi_->$ contains nearly pure atomic excitation. When the weak control field with $\Omega << \sqrt{N} g$ is coupled to the atoms from free space and tuned to be resonant with either one of the two normal modes ($\Delta = \lambda_+$ or $\lambda_-$), it creates two polariton excitation paths, and the resulting destructive quantum interference between the two paths suppresses the polarion (normal mode) excitations [20]. Therefore, optical switching of the cavity-atom polaritons can be performed by turning on or off of the control light. In particular, with $\Delta_c >> \sqrt{N} g$, the weak control laser can switch on and off of the polariton (normal mode) $|\varphi_+>$ (with nearly pure photonic excitation). Thus, the proposed scheme can function as a near perfect light switch in which photons from the cavity mode can be switched on or off by a free-space control laser with different frequency and at ultra-low intensities.

We treat the cavity field classically and derive the atomic susceptibilities in a semiclassical analysis in which the atoms are coupled by a free space control laser and a weak, intra-cavity probe laser. The susceptibility of the three-level $\Lambda$-type atomic medium at the frequency of the intra-cavity probe laser $\nu_p$, $\chi(\nu_p) = \chi'(\nu_p) + i\chi''(\nu_p)$, is derived by solving Schrodinger equations of the coupled atomic system under the condition $\rho_{aa} \approx 1$ (the intra-cavity probe field is much weaker than the coupling field so the atomic population is concentrated in the ground state $|a>$) and is given by

$$\chi(\nu_p) = \chi' + i\chi'' = \frac{K(\Delta_p - \Delta + i\gamma_{ab})}{|\Omega|^2 - (\Delta_p + i\Gamma/2)(\Delta_p - \Delta + i\gamma_{ab})}. \quad (1)$$

Here $K=n|\mu_{ea}|^2/\hbar\varepsilon_0$ (n is the atomic density), $\Gamma$ is the decay rate of the excited state |e> and $\gamma_{ab}$ is the decoherence rate between ground states |a> and |b>. The intensity of the cavity-transmitted probe field is

$$I_t(\upsilon_p) = \frac{I_{in}(\nu_p)(1-R)^2 \exp(-2k\chi''\ell)}{1+R^2\cdot\exp(-4k\chi''\ell)-2R\cdot\exp(-2k\chi''\ell)\cos(2ik(L-\ell+\chi'\ell))} \quad (2).$$

Here $\ell$ is the atomic medium length in the cavity mode. If the control laser is absent, the susceptibilities are linear and given by $\chi = \frac{K(i\Gamma/2-\Delta_p)}{\Delta_p^2+(\Gamma/2)^2}$. The transmission peak of the intra-cavity probe light occurs at the probe frequency at which the probe phase shift from the atoms, $2k\chi'\ell$, and the probe phase shift from the empty cavity, $2k(L-\ell)=\frac{2\pi(\Delta_p-\Delta_c)}{c/2L}$ cancel each other ($2k(L-\ell)+2k\chi'\ell=2m\pi$, m is an integer) [21]. Two such probe frequencies are derived at $\Delta_p=\frac{\Delta_c}{2}\pm\sqrt{\frac{\Delta_c^2}{4}+\frac{k\ell cK}{2\pi L}}$ and represent two cavity transmission peaks (assuming $\Delta_p^2 >> (\Gamma/2)^2$ and $\Gamma_c^2$, $\Gamma_c$ is the cavity linewidth). The number of atoms inside the intra-cavity probe beam is N=nA$\ell$=ODA/$\sigma_{ea}$ (OD is the medium optical depth, A is the cross section of the intra-cavity probe beam, and $\sigma_{ea}$ is the atomic cross section) and the cavity mode volume is V=AL. Then the two transmission peaks occur at

$$\Delta_p=\frac{\Delta_c}{2}\pm\sqrt{\frac{\Delta_c^2}{4}+\frac{k\ell cK}{2\pi L}}=\frac{\Delta_c}{2}\pm\sqrt{\frac{\Delta_c^2}{4}+\frac{|\mu_{ea}|^2\nu_p n\ell}{2\hbar\varepsilon_0 L}}=\frac{\Delta_c}{2}\pm\sqrt{\frac{\Delta_c^2}{4}+g^2N}=\lambda_\pm, \quad (3)$$

where $g^2N=\frac{|\mu_{ea}|^2\nu_p n\ell}{2\hbar\varepsilon_0 L}=G^2$. This result agrees with the previous derived eigenvalues of the first excited states of the coupled cavity-atom system. In Cavity QED, the multi-atom vacuum Rabi splitting (the normal mode splitting) is given by 2G=2$\sqrt{N}g$ (with $4Ng^2 >> \Gamma^2$ and $\Gamma_c^2$) [4-6].

Thus, the semiclassical analysis gives the identical result for the vacuum Rabi splitting as the QED analysis and the two transmission peaks at $\Delta_p = \lambda_\pm$ correspond to the two polariton resonances, or the two excited normal modes.

When the control laser is present, the atomic coherence and interference is induced in the coupled cavity-atom system. It is instructive to exam the dependence of $\chi$ on the control laser detuning $\Delta$ near the polariton resonance while the probe laser frequency is locked to the polariton resonance at $\Delta_p = \lambda_\pm$. Writing $\Delta = \Delta_p + \delta = \lambda_\pm + \delta$, $\chi$ in Eq.(1) is given by

$$\chi = \chi' + i\chi'' = K \frac{\delta |\Omega|^2 + \Delta_p \delta^2 + \gamma_{ab}^2 \Delta_p}{(|\Omega|^2 + \Delta_p \delta + \gamma_{ab}\Gamma)^2 + (\Delta_p \gamma_{ab} - \delta\Gamma)^2} + iK \frac{\gamma_{ab} |\Omega|^2 + \Gamma \delta^2 + \gamma_{ab}^2 \Gamma}{(|\Omega|^2 + \Delta_p \delta + \gamma_{ab}\Gamma)^2 + (\Delta_p \gamma_{ab} - \delta\Gamma)^2} \quad . \quad (4)$$

$\chi'$ represents the dispersive response of the atomic medium and contributes to the phase shift of the intra-cavity light field (when $\delta \neq 0$, the system can be used to obtain a large cross-phase modulation of the cavity-atom polaritons [22]). $\chi''$ represents the absorptive response of the atomic medium and leads to attenuation of the intra-cavity light field. The nonlinear absorptive susceptibility $\chi''_{non}$ induced by the control laser is

$$\chi''_{non} = \chi'' - \chi''(\Omega = 0) = K \frac{\gamma_{ab} |\Omega|^2 + \delta^2 \Gamma/2 + \gamma_{ab}^2 \Gamma/2}{(|\Omega|^2 + \Delta_p \delta + \gamma_{ab}\Gamma/2)^2 + (\Delta_p \gamma_{ab} - \delta\Gamma/2)^2} - K \frac{\Gamma/2}{\Delta_p^2 + (\Gamma/2)^2} \quad . \quad (5)$$

At $\delta = 0$ where the optical switching operates,

$$\chi''_{non} = K \frac{|\Omega|^2 (\gamma_{ab} \Delta_p^2 - \gamma_{ab}\Gamma/2 - |\Omega|^2 \Gamma/2)}{((|\Omega|^2 + \gamma_{ab}\Gamma/2)^2 + (\Delta_p \gamma_{ab})^2)(\Delta_p^2 + (\Gamma/2)^2)} \quad . \quad (6)$$

Under conditions of the strong cavity-atom coupling ($4Ng^2 >> \Gamma^2$ and $\Gamma_c^2$) and a weak control field ($|\Omega|^2 << \Delta_p^2 \gamma_{ab}/\Gamma$), the nonlinear absorptive susceptibility becomes

$$\chi''_{non} = K \frac{|\Omega|^2 \gamma_{ab}}{(|\Omega|^2 + \gamma_{ab}\Gamma/2)^2 + (\Delta_p \gamma_{ab})^2} \quad .$$ The intensity absorption length of the intra-cavity light

field for a single pass in the cavity is $2k\chi''_{non}\ell$. With the cavity feedback, the cavity-enhanced intensity absorption length is $2k\chi''_{non}\ell\frac{2F}{\pi}$ where $F = \frac{\pi\sqrt{R}}{1-R}$ is the cavity finesse. If we define the critical level of the switching off of the polariton at the intensity attenuation of $e^{-1}$ of the full polariton intensity, then by setting $2k\chi''_{non}\ell\frac{2F}{\pi} = 1$, we derive the minimum value $\Omega_{min}$ for the control laser required to perform the polariton switching in the cavity QED system,

$$|\Omega_{min}| = \frac{\Delta_p}{2g\sqrt{N}}\sqrt{\gamma_{ab}\Gamma_c} \ . \qquad (7)$$

Then, the required minimum switching intensity of the control laser is $I_{min} = c\varepsilon_0 E^2 = c\varepsilon_0(\hbar\Omega_{min}/\mu_{be})^2$. As examples, we consider two separate cases for cold Rb atoms confined in a cavity with L=5cm. (1) When the empty cavity is tuned to the atomic transition frequency ($\Delta_c$=0), then $\Delta_p = \sqrt{N}g$ and the polariton consists of equal (50%) superposition of the atomic excitation and photonic excitation. With R=0.98, $\sqrt{N}g = 10\Gamma \gg \Gamma$, and $\gamma_{ab}$=0.01$\Gamma$ (a conservative $\gamma_{ab}$ value since the life time of the ground state coherence as long as a few ms has been observed in an experiment with cold Rb atoms [23], which corresponds to $\gamma_{ab}$~$10^{-4}\Gamma$ ), Eq. (7) gives $I_{min}$=0.04 mw/cm$^2$ ($\Omega_{min}$=0.089$\Gamma$). (2) When the empty cavity is tuned far away from the atomic transition frequency ($\Delta_c$=50$\Gamma$), with R=0.99, $\gamma_{ab}$=0.01$\Gamma$, and $\sqrt{N}g = 10\Gamma$, Eq. (7) gives $I_{min}$=0.45 mw/cm$^2$ ($\Omega_{min}$=0.3$\Gamma$). Note that in case 2, the polariton $|\varphi_+\rangle$ contains ~ 98% of the photonic (photon) excitation. Switching such polaritons practically can be viewed as switching pure photons. The required control light intensity for the polariton switching is greater in case (2) because of the large cavity detuning $\Delta_c$. But in both cases, $I_{min}$ is well below the saturation intensity of the Rb transition ($\approx$1.6 mw/cm$^2$). For a single photon (795 nm) in a 1 μs pulse and focused to a spot size of the atomic cross section $\sigma_{ea} \approx 10^{-9}$ cm$^2$, the photon intensity is 0.23 mW/cm$^2$. Therefore, the proposed system is capable of performing optical switching of the cavity-atom polaritons at single photon levels of the control light field.

Next, we present numerical calculations from Eq. (1) and (2) for a practical cavity QED system consisting of cold Rb atoms confined in a moderate size cavity (L=5 cm, $\ell$=1 mm, and OD=$n\sigma_{ea}\ell$ ~ 4). First, we discuss the case for the polariton switching with the equal weight of the atomic excitation and the photonic excitation (the empty cavity is tuned to the atomic transition frequency, $\Delta_c = \nu_c - \nu_{ea}$=0). Fig. 2 plots the transmitted light intensity I$_t$ (normalized to the input light intensity I$_{in}$) versus the probe frequency detuning $\Delta_p/\Gamma$ in (a) without the control laser and (b) with the control laser. Without the control laser, the transmission spectrum exhibits two symmetrical peaks located at $\Delta_p = \pm\sqrt{N}g$ , corresponding to the two excited polaritons (the normal modes). When the control laser is present and tuned in frequency to one of the polariton resonance ($\Delta_c = \sqrt{N}g$ in Fig. 2(b)), the polariton excitation is suppressed at $\Delta_p = \sqrt{N}g$ by the destructive interference induced by the control laser [20]. In order to see the effect of the ground state decoherence on the polariton switching, we plot the transmitted light intensity I$_t$/I$_{in}$ versus the control laser detuning $\delta/\Gamma$ for various $\gamma_{ab}$ values in Fig. 3(a). The calculations show that as $\gamma_{ab}$ increases, the transmitted light intensity also increases, which reduces the polariton switching efficiency. The linewidth of the dark dip induced by the destructive quantum interference at the polariton resonance increases with the decoherence rate $\gamma_{ab}$. Our calculation shows that the dip linewidth is ultimately determined by $\gamma_{ab}$ and is power broadended by the control laser when $\Omega > \gamma_{ab}$. Fig. 3(b) plots the transmitted light intensity I$_t$/I$_{in}$ at the polariton resonance ($\delta$=0 or $\Delta_p=\Delta=\sqrt{N}g$ ) versus the Rabi frequency of the control laser. The required $\Omega$ value for switching off the polaritons depends on the decoherence rate $\gamma_{ab}$ and increases with the increasing $\gamma_{ab}$ value. If we define the switching off at $I_t(\Omega_{min})/I_t(\Omega=0)=e^{-1}$, then $\Omega_{min}$ derived from Fig. 3(b) is consistent with Eq. (7).

Calculations of the cavity-atom polariton switching under the condition of a large empty-cavity detuning ($\Delta_c=-50\Gamma$) are presented in Fig. 4 and Fig. 5. Fig. 4 plots $I_t/I_{in}$ versus the probe detuning $\Delta_p/\Gamma$ in (a) without the control laser and (b) with the control laser. Without the control laser, the intensity transmission spectrum is asymmetrical with a large peak at $\lambda_+\approx\Delta_c$, representing the polariton with nearly pure photonic excitations ($|\varphi_+>\approx|a>|1>$, the peak transmission $I_t/I_{in}\approx96\%$). There is also a small peak at $\lambda_- = -\dfrac{g^2N}{\Delta_c}$, representing the polariton with nearly pure atomic excitations ($|\varphi_->\approx|e>|0>$). When the control laser is present and tuned to the polariton resonance at $\Delta=\lambda_+\approx\Delta_c$, the polariton excitation is suppressed at the resonance ($\Delta_p=\lambda_+\approx\Delta_c$) by the destructive interference induced by the control laser. Again, the ground state decoherence reduces the coherence and interference, and its effect on the polariton switching is revealed in Fig. 5(a) which plots the transmitted light intensity $I_t/I_{in}$ versus the control laser detuning $\delta/\Gamma$ for a series of $\gamma_{ab}$ values. The calculations show that as the decoherence rate $\gamma_{ab}$ increases, the transmitted light intensity increases, which reduces the polariton switching efficiency. The linewidth of the dark dip induced by the destructive quantum interference at the polariton resonance ($\delta=0$ or $\Delta=\lambda_+\approx\Delta_c$) is limited by the decoherence rate $\gamma_{ab}$. Fig. 5(b) plots the transmitted light intensity $I_t/I_{in}$ at the polariton resonance ($\Delta_p=\Delta=\lambda_+\approx\Delta_c$) versus the Rabi frequency of the control laser. The required $\Omega$ value for the effective photon switching increases with the increasing $\gamma_{ab}$ value and is greater than $\Omega$ values obtained under the condition of $\Delta_c=0$. Again, if we define the switching off at $I_t(\Omega_{min})/I_t(\Omega=0)= e^{-1}$, then $\Omega_{min}$ derived from Fig. 5(b) is consistent with Eq. (7).

Fig. 6 plots the transmitted intensity $I_t/I_{in}$ versus the control laser detuning $\delta/\Gamma$ and the control Rabi frequency $\Omega$ for (a) $\Delta_c=0$ and (b) $\Delta_c=-50\Gamma$. As $\Omega$ increases, the transmitted intensity decreases (down to zero for a sufficiently large $\Omega$) and the dip linewidth increases. With a far-detuned cavity ($\Delta_c \gg \Gamma$), the power broadening of the dip linewidth occurs at higher $\Omega$ values than that with a resonant cavity ($\Delta_c=0$). The required energy density of the control laser for switching nearly pure cavity photons (at $\Delta_c \gg \Gamma$) is higher than that required for switching of the cavity-atom polaritons with the equal mixture of the atomic excitation and photonic excitation (at $\Delta_c=0$). The control light is able to perform the cavity-atom polariton switching at ultra-low intensities because of the narrow linewidth of the interference dip. However, the small linewidth of the interference dip in the cavity-atom system implies a long switching time and correspondingly, a slow switching speed. If the power broadening is negligible, the switching time in the cavity-atom system is given by $1/\gamma_{ab}$ (~20 μs for $\gamma_{ab}=0.01\Gamma$ with $\Gamma=6$ MHz in the Rb atoms). When $\Omega \gg \gamma_{ab}$, the dip linewidth is power broadened, and correspondingly, the switching time is reduced.

In conclusion, we have proposed and analyzed a scheme for all optical switching of the cavity-atom polaritons in a multi-atom cavity QED system. The analysis shows that efficient switching of cavity-atom polaritons with an ultra-low control light can be realized in a practical experimental system with cold Rb atoms confined in a moderate finesse cavity if the linewidth of the control and probe lasers is much smaller than the dip linewidth. When the cavity is detuned far from the atomic transition frequency, the polaritons consist of nearly pure photonic excitations and the scheme can then be used effectively for all optical switching of cavity photons at low light levels. The proposed scheme combines studies of cavity QED and nonlinear optics at ultra-low light levels, and may be useful for quantum electronics and photonics applications.

This work is supported by the National Science Foundation under Grant No. 0757984.

Figure Captions

Fig. 1 Three level atoms coupled to a cavity field and a free-space control field.

Fig. 2 $I_t / I_{in}$ ($I_t$ is the cavity transmitted probe intensity and $I_{in}$ is the input probe intensity) versus the normalized probe frequency detuning $\Delta_p/ \Gamma$. (a) Without the control laser. (b) With the control laser ($\Delta= \sqrt{N} g$ and $\Omega=0.2\Gamma$). Parameters used in the calculations are: $\gamma= 0.01\Gamma$, optical depth $n\sigma_{ea}\ell=4$, L=5 cm, $\Delta_c=0$, and R=0.98. The inset Fig. plots the expanded view of the transmitted probe intensity between $\Delta_p/ \Gamma$= 9 to 13.

Fig. 3 (a) the transmitted light intensity $I_t / I_{in}$ versus the normalized control frequency detuning $\delta/ \Gamma$ ($\delta=\Delta - \sqrt{N} g$ ) and (b) $I_t / I_{in}$ versus the control Rabi frequency $\Omega/ \Gamma$ for a series of $\gamma_{ab}$ values. The other parameters are the same as those in Fig. 2.

Fig. 4 $I_t / I_{in}$ versus the normalized probe frequency detuning $\Delta_p/ \Gamma$. (a) Without the control laser. (b) With the control laser ($\Delta=\lambda_+\approx\Delta_c$ and $\Omega=0.2\Gamma$). Parameters used in the calculations are: $\gamma= 0.01\Gamma$, optical depth $n\sigma_{ea}\ell=4$, L=5 cm, $\Delta_c=50\Gamma$, and R=0.99. The inset Fig. plots the expanded view of the transmitted probe intensity between $\Delta_p/ \Gamma$= 52.8 to 53.2.

Fig. 5 (a) $I_t / I_{in}$ versus the normalized control frequency detuning $\delta/ \Gamma$ ($\delta=\Delta - \sqrt{N} g$ ) and (b) $I_t / I_{in}$ versus the control Rabi frequency $\Omega/ \Gamma$ for a series of $\gamma_{ab}$ values. The other parameters are the same as those in Fig. 4.

Fig. 6 $I_t / I_{in}$ versus $\Omega/\Gamma$ and $\delta/ \Gamma$. In (a), R=0.98, $\Delta_c=0$, $\Delta_p= \sqrt{N} g$ , and $\delta=\Delta - \sqrt{N} g$ . In (b), R=0.99, $\Delta_c=-50\Gamma$, $\Delta_p=\lambda_+\approx\Delta_c$, and $\delta=\Delta - \Delta_p$. The other parameters are the same as those in Fig. 4

.

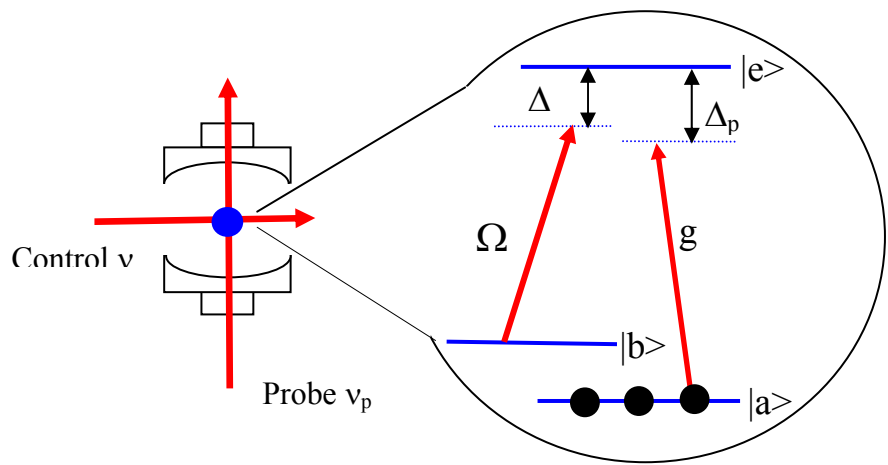

Fig. 1

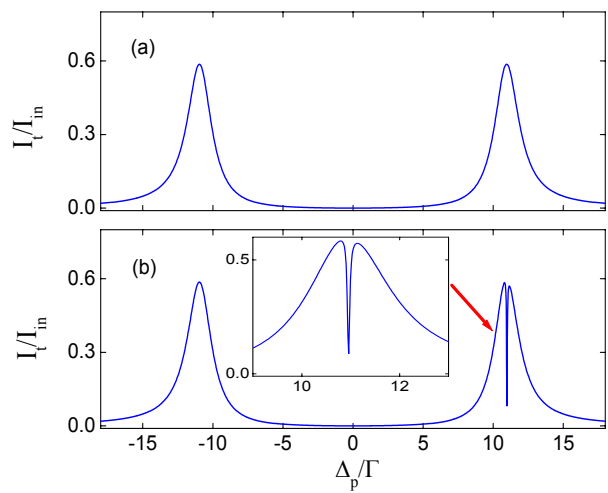

Fig. 2

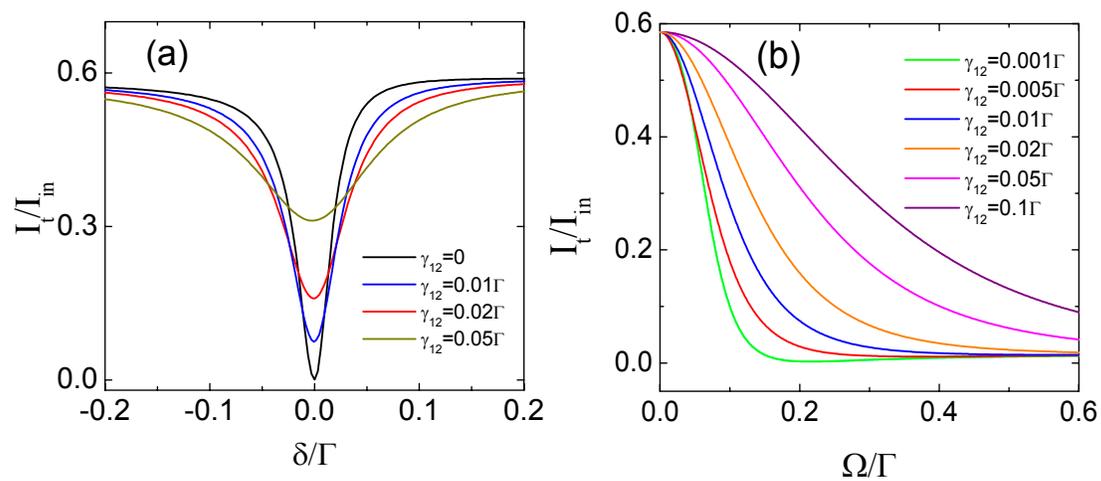

Fig. 3

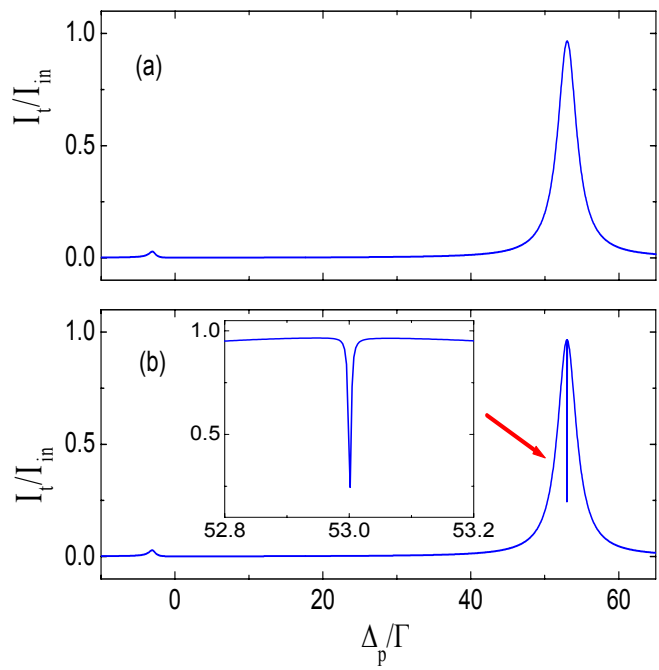

Fig. 4

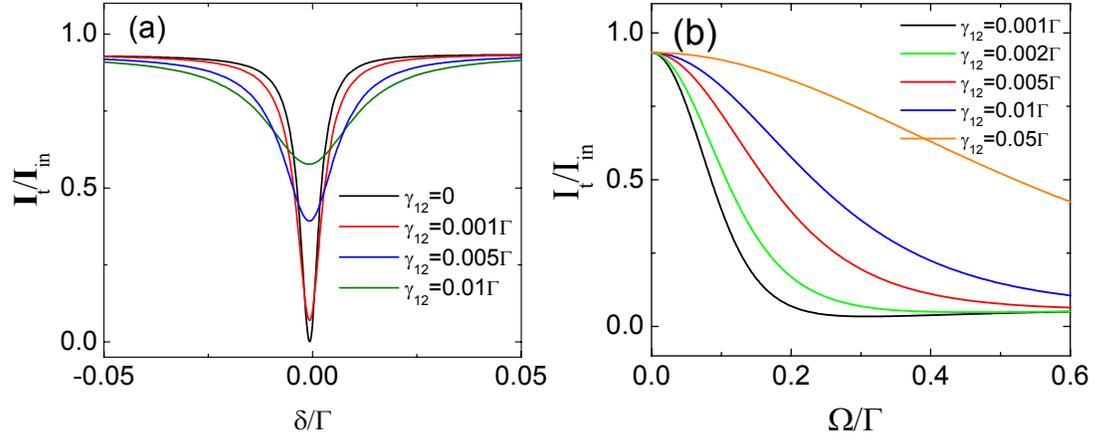

Fig. 5

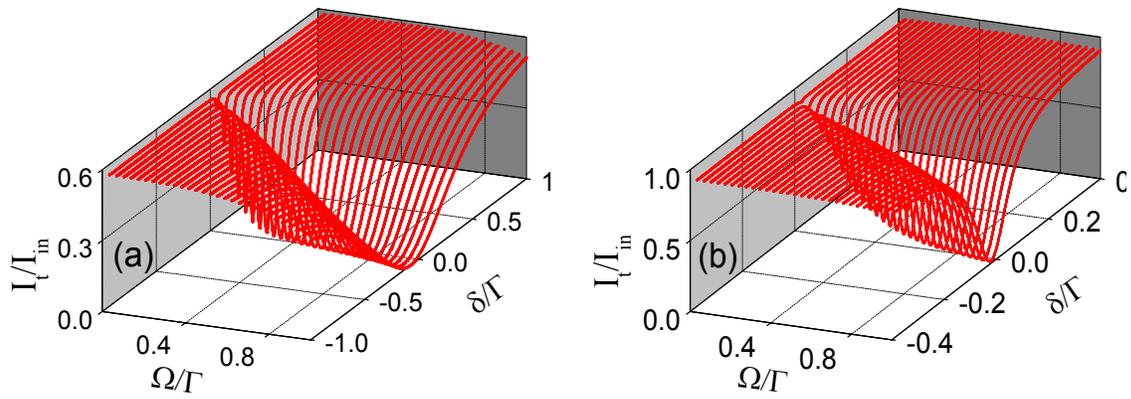

Fig. 6